\documentclass{WileyMSP-template}

\usepackage{graphicx}
\usepackage{dcolumn}
\usepackage{bm}
\usepackage{color}
\usepackage[utf8]{inputenc}
\usepackage[T1]{fontenc}
\usepackage{mathptmx}
\usepackage{amsmath,amssymb}
\usepackage{siunitx}
\usepackage[version=4]{mhchem}
\usepackage[normalem]{ulem}
\usepackage[euler]{textgreek}
\usepackage{xfrac}

\newcommand{\SMU}{17~\textmu m }

\DeclareGraphicsExtensions{.pdf}

\begin{document}

\pagestyle{fancy}
\rhead{\includegraphics[width=2.5cm]{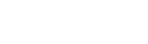}}

\title{Demonstration and frequency noise characterization of a $17$~\textmu m quantum cascade laser}
\maketitle


\author{Mathieu Manceau,}
\author{Thomas E. Wall,}
\author{Hadrien Philip,}
\author{Alexei Baranov,}
\author{Olivier Lopez,}
\author{Michael R. Tarbutt,}
\author{Roland Teissier and}
\author{Benoît Darquié*}

\affiliations{
M. Manceau, O. Lopez, B. Darquié\\
Laboratoire de Physique des Lasers, CNRS, Université Sorbonne Paris Nord, Villetaneuse, France \\
Email : benoit.darquie@univ-paris13.fr

T.E. Wall\\
Present address: Space Science and Technology Department, STFC Rutherford Appleton Laboratory, Harwell Campus, Didcot OX11 0QX, UK\\

T.E. Wall, M.R. Tarbutt\\
Centre for Cold Matter, Blackett Laboratory, Imperial College London, Prince Consort Road, London SW7 2AZ, United Kingdom, \\

H. Philip, A. Baranov, R. Teissier\\
IES, University of Montpellier, CNRS, 34095 Montpellier, France}%


\begin{abstract} 

We evaluate the spectral performance of a novel continuous-wave room-temperature distributed feedback quantum cascade laser operating at the long wavelength of $17$~\textmu m. By demonstrating broadband laser absorption spectroscopy of the $\nu_2$ fundamental vibrational mode of \ce{N2O} molecules, we have determined the spectral range and established the spectroscopic potential of this laser. We have characterized the frequency noise and measured the line width of this new device, uncovering a discrepancy with the current consensus on the theoretical modeling of quantum cascade lasers. Our results confirm the potential of such novel narrow-line-width sources for vibrational spectroscopy. Extending laser spectroscopy to longer wavelength is a fascinating prospect that paves the way for a wide range of opportunities from chemical detection, to frequency metrology as well as for exploring light-matter interaction with an extended variety of molecules, from ultra-cold diatomic species to increasingly complex molecular systems.

\end{abstract}


\section{Introduction}

\begin{figure}[t]
\centering \includegraphics[scale=0.45]{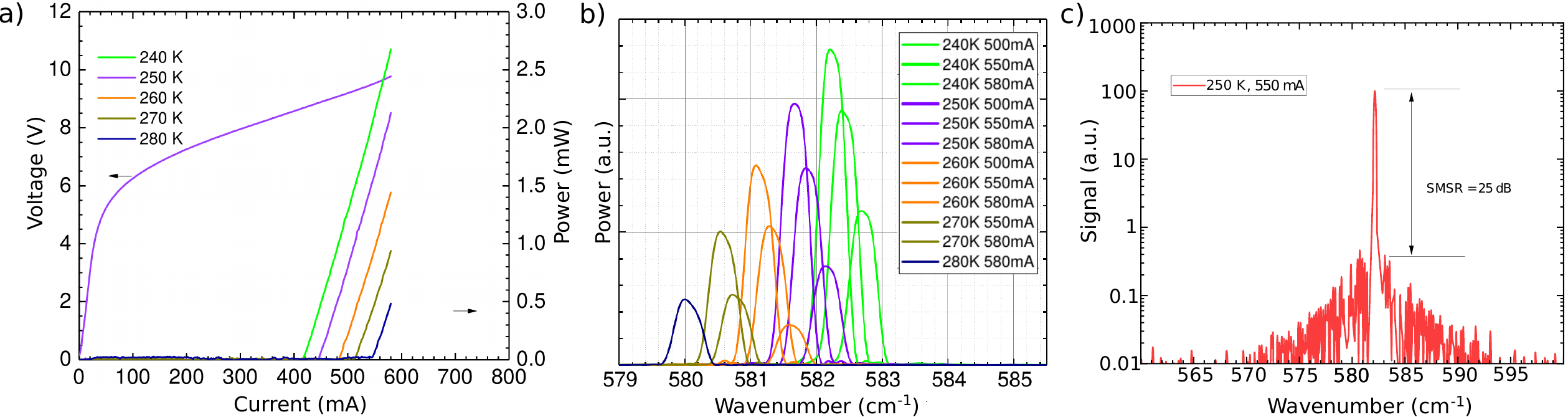} 
\caption{\label{fig:QCLCharacteristics} a) Optical output power and voltage \emph{versus} drive current recorded for various QCL temperatures. b) Emission spectra, measured by a Fourier transform infrared (FTIR) spectrometer  for various QCL temperatures and drive currents shown in the legend. c) Emission spectrum measured at 250~K and 550~mA at higher FTIR spectrometer resolution, demonstrating a Side Mode Suppression Ratio (\textbf{SMSR}) of at least 25~dB.
} \end{figure}

Stimulated by the invention of the quantum cascade laser (QCL)~\cite{faist1994}, applications relying on mid-infrared (MIR) radiation have progressed at a very rapid pace in recent years. These range from free-space optical communications~\cite{corrigan2009,liu_mid_2019,corrias2022}, gas sensing~\cite{Nikodem_2012_d, Daghestani_2014, Martin-Mateos_2017, Robinson_2021, kuenning2024} and trace detection~\cite{galli_spectroscopic_2016}, and high-resolution spectroscopy~\cite{bielsa_hcooh_2008, asselin_characterising_2017,borri2019,Dambrosio2019,arellano_probing_2024}, to metrology and frequency referencing~\cite{bielsa_2007,sow2014,hansen_quantum_2015,argence2015,insero2017,santagata2019, chomet_highly_2023, tran_near-_2024, chomet_heterodyne_2024,tran2025}, as well as fundamental physics measurements~\cite{Mejri2015,cournol2019,lukusa_mudiayi_linear_2021}. Unlike MIR gas lasers, such as CO and \ce{CO2} lasers, QCLs can provide broad and continuous frequency tuning over several hundreds of gigahertz. QCLs are also compact, robust and low-power devices compared with other more complex MIR sources based on frequency down-conversion \cite{petrov2012, schliesser2012}, such as optical parametric generators (OPG), oscillators (OPO) \cite{leindecker2011}, or difference frequency generators (DFG) \cite{sotor2018,lamperti2020}, rendering QCLs better suited for field deployment. Moreover, MIR QCLs can be easily interchanged and available wavelengths cover large parts of the MIR region from 2.6~\cite{cathabard_quantum_2010} to 28~\textmu m~\cite{ohtani_far-infrared_2016}, which is not the case with most of the other MIR sources. Fiber \cite{jackson2020} or crystalline (Cr:ZnSe, Tb:\ce{KPb2Cl5})  MIR  lasers \cite{vasilyev2016} are, for example, limited to the 2 to 5~\textmu m~spectral region. 

Distributed feedback (DFB) QCLs~\cite{faist1997}, which have a grating embedded in the laser cavity, are single longitudinal mode narrow-band lasers, and thus a solution of choice for high-resolution molecular spectroscopy. However, until recently continuous wave (CW) DFB QCLs operating at room-temperature were only available in the 4 to 11~\textmu m window.
Extending such technologies to longer wavelengths is important for a range of  applications: (i) strong vibrational signatures of small hydrocarbons (such as ethene, ethane, acetylene, propane), of larger aromatics (such as BTEX – benzene, toluene, ethylbenzene, and xylenes), of nitrous oxide and uranium hexafluoride are found in the 12-18~\textmu m spectral window~\cite{pirali2006,huang_temperature-insensitive_2011,fuchs_distributed_2011,lamperti2020,karhu2020,ayache2022,elkhazraji2023high,elkhazraji2023,wang_wavelength_2025,kinjalk2024}; in particular, it hosts the strongest absorption lines of C$_2$H$_2$, BTEX and UF$_6$; (ii) long wavelength (N and Q astronomical bands) QCLs would be valuable in radio-astronomy as local oscillators in heterodyne detectors \cite{bourdarot2020,bourdarot2021}; 
(iii) long wavelength QCLs can be used to develop a new set of MIR frequency standards in the wavelength range beyond 15 \textmu m based on the vibrational frequencies of trapped, ultracold molecules such as SrF, CaF, YbF, BaF and YO~\cite{wang_wavelength_2025,barontini_measuring_2022}
(iv) the increase in the  number of vibrational modes in large polyatomic molecules leads to intramolecular vibrational redistribution (IVR)\cite{nesbitt1996} or other more subtle rovibrational coupling mechanisms~\cite{liu_ergodicity_2023} which results in severe spectral fractionation and/or broadening; bringing increasingly complex molecular systems within reach of precise spectroscopic measurements offers promising perspectives in astrophysics, earth sciences, quantum technologies, metrology and fundamental physics, but requires working at increasingly low transition energies at which intramolecular vibrational couplings are correspondingly reduced \cite{brumfield2012,spaun2016,liu_ergodicity_2023}. In this paper, we report the characterization and operation of the longest wavelength room-temperature CW DFB QCL technology~\cite{nguyen2019}. Our source has been designed to operate at a wavelength of 17.2~\textmu m, in coincidence with the vibration of calcium monofluoride (CaF), as it could constitute an enabling technology for a MIR frequency standard based on ultracold CaF samples~\cite{wang_wavelength_2025,barontini_measuring_2022}.

The article is structured as follows: we describe the spectroscopy of the \emph{v}$_2$ mode of \ce{N2O} using the QCL, the first absorption spectroscopy reported at \SMU wavelength using a QCL. We then present the measurement of the QCL frequency noise, using an \ce{N2O} absorption line as a frequency discriminator. Finally, applications of the QCL are described, in particular the precise spectroscopy of ultra-cold molecules.

\section{17 \textmu m QCL spectroscopy}

The active region of the QCL is formed from InAs/AlSb \cite{baranov2016,nguyen2019}. The laser chip is mounted on a Peltier-cooled module. Laser characteristics are shown in \textbf{Figure~\ref{fig:QCLCharacteristics}}, demonstrating that several milliwatts of optical power can be generated over a frequency range of $\sim3~\textrm{cm}^{-1}$ ($\sim 100$~GHz) by tuning the temperature and drive current. The QCL beam is collimated by a parabolic mirror mounted a few millimeters away from the laser chip.

\begin{figure*}[t]
\centering \includegraphics[scale=.70]{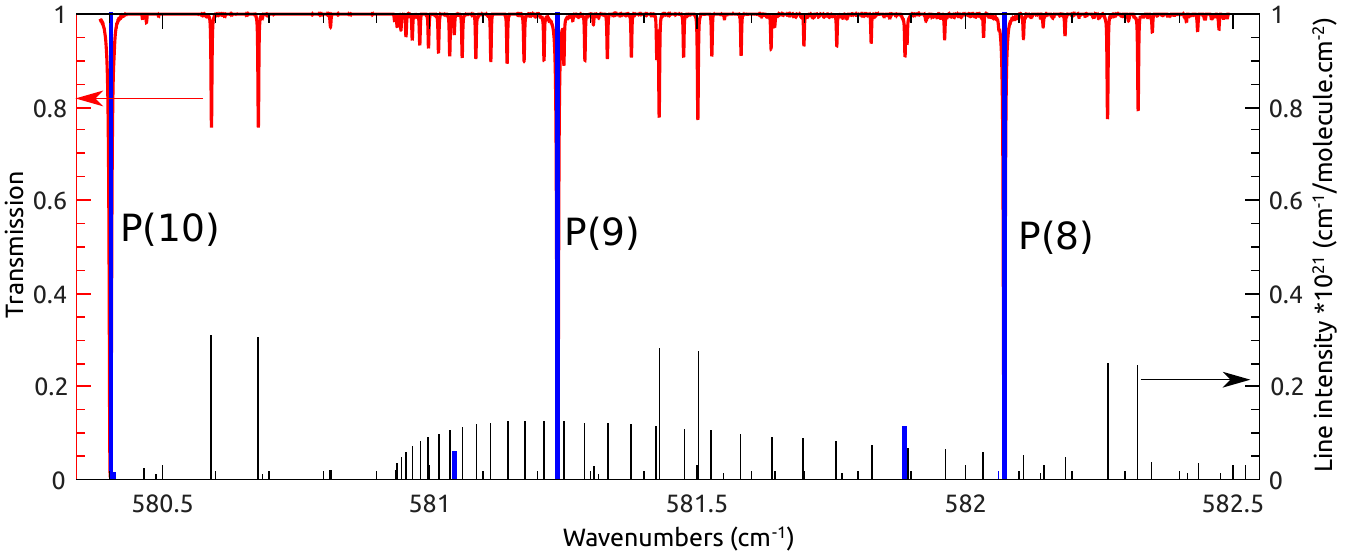}
\caption{\label{fig:N2OSpectra} Measured linear absorption spectrum of \ce{N2O} (red curve)
at a pressure of 500~Pa. The bar spectrum indicates the line frequencies and intensities reported in  the NIST database~\cite{makiOnline98} for five different \ce{N2O} isotopologues, the main contributions coming from $^{14}$N$_2^{16}$O (99\% abundance). The blue sticks are characterized by a better accuracy than the black sticks~\cite{makiOnline98}.  
}
\end{figure*}

 We have carried out \ce{N2O} absorption spectroscopy over the full tuning range of the QCL. There has been very little laser spectroscopy of \ce{N2O} around 17~\textmu m so far. The only previous \ce{N2O} laser spectroscopy in this spectral region was performed in 1979 by Reisfeld and Flicker~\cite{reisfeld:79} and in the 1990s by Baldacchini and co-workers~\cite{Baldacchini:92,Baldacchini:93} who both used a lead salt laser. These works were hampered by the poor performance of the sources. Lead salt lasers exhibit broad emission line widths, are affected by mode hopping and their spectral properties vary after temperature cycling. QCLs offer a much more reliable and precise alternative as we demonstrate in this article and in an accompanying paper that focuses on the spectroscopy of \ce{N2O}~\cite{wang_wavelength_2025}.

In order to explore the entire spectral coverage of the QCL, seven spectra have been recorded, each at different laser operating conditions. For these measurements, the laser temperature set-point was varied from -28.5\textdegree C to -2.6\textdegree C. At each temperature, a 100~ms-period current ramp is applied to the QCL around a mean laser drive current (\emph{e.g.} $\sim \pm 50$~mA around $\sim 525$~mA at low temperatures) allowing the frequency to be swept over a span of a few gigahertz to $\sim 20$~GHz depending on the temperature. \ce{N2O} absorption data spanning the 580.4~cm$^{-1}$ to 582.5~cm$^{-1}$ window have been recorded at a pressure of 500~Pa in a 24~cm-long gas cell using a liquid-nitrogen-cooled photoconductive HgCdTe (MCT) detector. Absorption spectra are always `contaminated' by unwanted background fringes resulting from the interference between the transmitted laser beam and parasitic reflections. To eliminate this spurious signal and convert raw data into absolute absorption spectra, the following measurements are recorded: (i) the laser beam is blocked and a frequency scan is acquired, which allows the MCT dark signal to be recorded; (ii) the laser is unblocked, the gas cell is filled with 500~Pa of \ce{N2O}, and a frequency scan is recorded; (iii) the gas cell is evacuated and a frequency scan is recorded in order to measure the transmission of the gas cell itself. These reference spectra were used to produce absolute absorption spectra with parasitic fringes largely suppressed and dark signals subtracted. We calibrate the frequency axis using the NIST database~\cite{makiOnline98}. For each measurement, around eight strong absorption features, distributed across the entire spectrum, are selected as calibration lines. We fit a Gaussian function to each, find the respective QCL currents that produce the frequency of each Gaussian centroid, and associate each with the corresponding frequency from the NIST database. Since the line shape has minimal influence on the accuracy of the centroid determination, a simple Gaussian fit was used; the other details of the line shape are not essential here.
A third-order polynomial fit to the line centre data allows us to establish the relationship between the QCL current and the absolute frequency. Individual measurements were recorded and combined into a single spectrum, shown in \textbf{Figure~\ref{fig:N2OSpectra}}. The three intense features correspond to the $P(8)$, $P(9)$ and $P(10)$ lines of the (01$^1$0) -- (00$^0$0) fundamental band of $^{14}$N$_2^{16}$O $\nu_2$ bending mode. Most other features correspond to hot bands of the same mode, a few to the bending modes of other isotopologues.

\section{Frequency noise}

\begin{figure}[h]
\centering \includegraphics[scale=0.5]{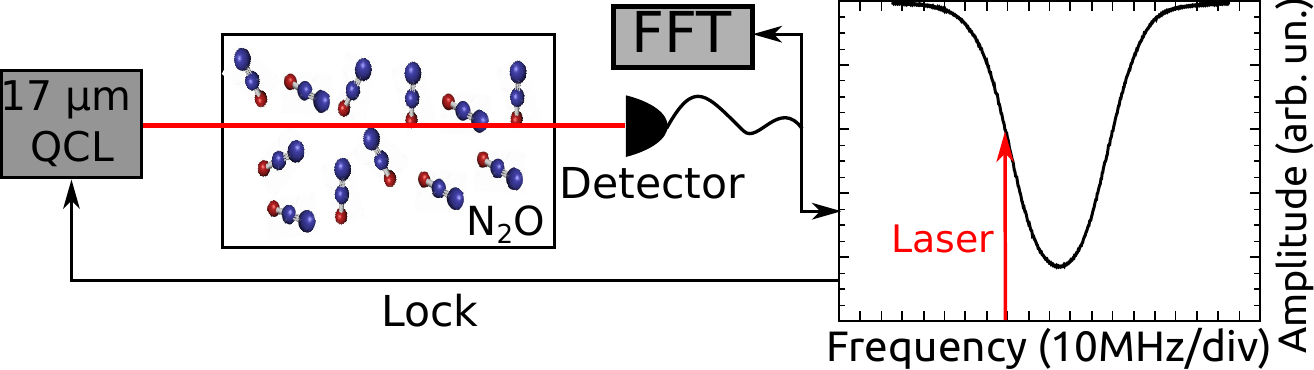} 
\caption{\label{fig:Setup}  QCL frequency noise measurement 
setup. A 24-cm long cell filled at a \ce{N2O} pressure of 80~Pa is used for the QCL frequency noise analysis. The graph shows a measurement of the $P(9)$ line of the $v_2$ vibrational bending mode at 581.24~cm$^{-1}$ used as as a frequency-to-amplitude converter. The arrow shows the QCL frequency when recording the frequency noise power spectral density. The QCL is loosely locked to the side of the molecular line. Intensity fluctuations proportional to the laser frequency noise are recorded on the detector and processed by the Fast Fourier Transform (FFT) spectrum analyzer. Both the gas cell and the MCT detector are tilted to minimize optical feedback to the QCL.}\end{figure}

As illustrated in \textbf{Figure~\ref{fig:Setup}}, we use the side of the $P(9)$ line at 581.24~cm$^{-1}$ as a frequency-to-amplitude converter to measure the frequency noise of the \SMU QCL~\cite{bartalini2010,bartalini2011,sow2014}. This measurement is carried out at a \ce{N2O} pressure of 80~Pa, resulting in a large absorption signal and limited collision-induced broadening, and the QCL is operated at a temperature of 258~K  and a current of 570~mA.

\begin{figure}[h]
\centering \includegraphics[scale=0.6]{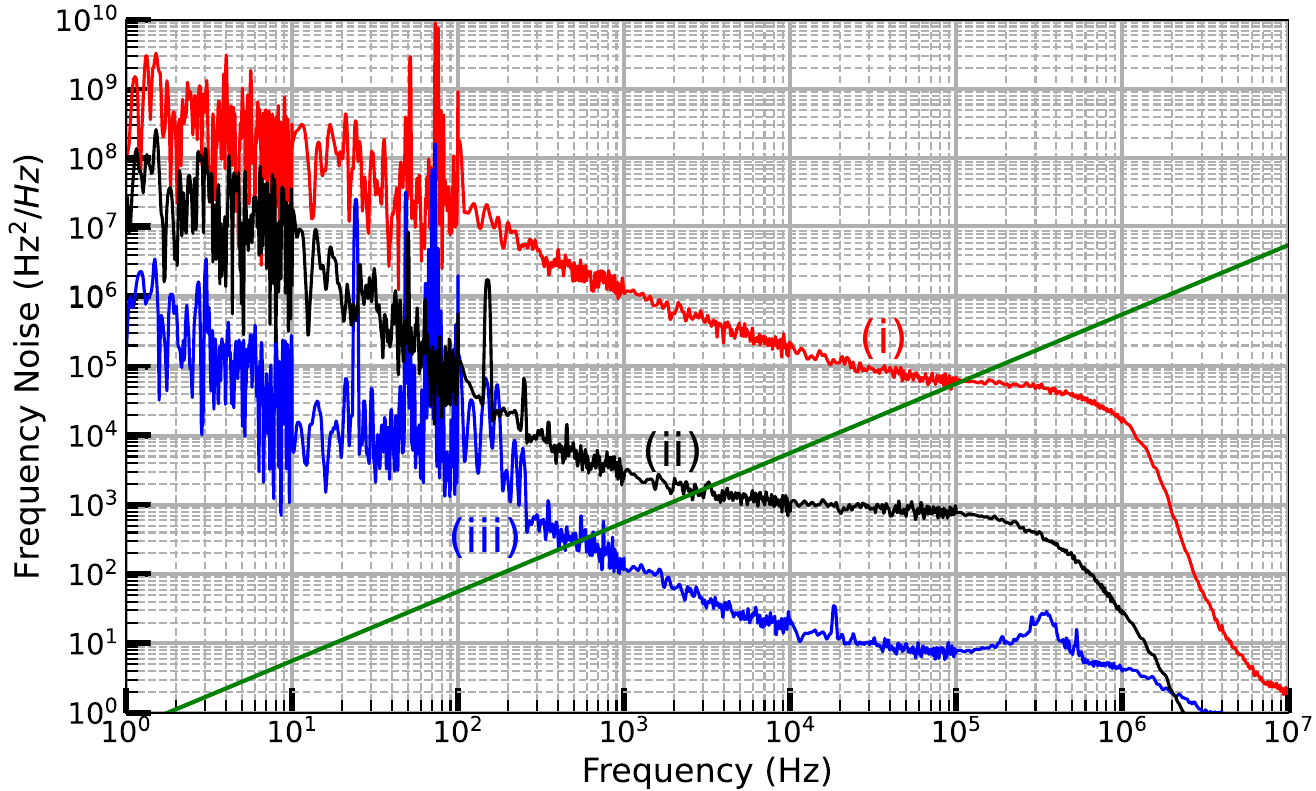}
\caption{\label{fig:NoisePSD} Frequency noise power spectral density (PSD) of the \SMU QCL (red line (i)). The contributions from the laser driver current noise (black line (ii)) and laser intensity noise (blue line (iii)) are also shown for comparison. The laser intensity contribution has been recorded with the same QCL parameters (temperature, current) as for the frequency noise measurement. The optical power hitting the detector is the same for both measurements. The green line is the $\beta$-separation line defined by $ 8 \ln(2) f / \pi^2$ and used to estimate the laser line width from the PSD measurement.}
\end{figure}

The \SMU QCL exhibits long-term frequency drifts on the order of a few hundred kilohertz per second, as determined through repeated measurements of the \ce{N2O} absorption line. To mitigate this drift and ensure the laser frequency remains on the absorption line during frequency noise measurements, the laser frequency is locked to the side of the $P(9)$ transition. The stabilization circuit has a bandwidth of approximately 1~Hz, allowing us to correct for slow frequency drifts without narrowing the laser's linewidth. The resulting amplitude noise generated by the molecular line used as a frequency discriminator is recorded on the MCT detector and processed by a Fast Fourier Transform (FFT) spectrum analyzer. The frequency-to-amplitude conversion coefficient,  the slope of the absorption profile at the position of the laser, see Figure~\ref{fig:Setup},  is obtained after recording the $P(9)$ line and fitting a Voigt profile to the corresponding data. Here the Doppler width is used to calibrate the frequency scale by fixing it to its value at 293~K, the temperature of the \ce{N2O} cell. The resulting full-width-at-half-maximum (FWHM) of $\sim46$~MHz comes from a combination of the Gaussian Doppler width, $\sim32.2$~MHz FWHM,  the Lorentzian pressure broadening width, $\sim5.1$~MHz FWHM, and the Beer-Lambert distortion at our $\sim83$\% level of absorption. This width is much larger than the QCL line width at any time scale (see below), which justifies considering the discriminator response as linear \cite{kashanian2016}. \textbf{Figure~\ref{fig:NoisePSD}} shows the resulting frequency noise power spectral density (PSD) of the QCL (red curve (i)). We have also evaluated potential contributions to the measured frequency noise resulting from the current noise of a home-made low-noise current source (black curve (ii)) and from laser intensity noise (blue curve (iii)). The contribution from the current source was calculated by multiplying its measured current noise spectrum by the QCL’s DC current-to-frequency tuning coefficient at 260~K and 570~mA ($\sim$240~MHz/mA, as determined from the FTIR spectral measurements shown in \textbf{Figure~\ref{fig:QCLCharacteristics}b}).  The measured amplitude noise when the laser is positioned on the side of the molecular absorption line is not only a consequence of frequency noise; it is also affected by pure laser intensity noise. To isolate this contribution, the laser is tuned far from any molecular resonance—i.e., where frequency-to-amplitude conversion via molecular absorption does not occur. The pure laser intensity noise is then measured, and its contribution to the frequency noise PSD is obtained by using the frequency-to-amplitude coefficient used to calculate curve (i). The result is shown as the blue curve (iii) in \textbf{Figure~\ref{fig:NoisePSD}}. Both the laser intensity noise and the current source noise contributions are thus negligible. Therefore, the red curve in \textbf{Figure~\ref{fig:NoisePSD}} represents the free-running frequency noise PSD of the QCL.

As seen in \textbf{Figure~\ref{fig:NoisePSD}}, at low frequencies ($\lesssim 100$~ kHz), the QCL frequency noise is dominated by the usual $1/f$ flicker noise. For frequencies greater than 100~kHz, a noise plateau appears, as has been reported for other QCLs at shorter wavelengths \cite{bartalini2010,bartalini2011}. This white noise level $N_{\mathrm {w}} \simeq 60\times10^3$~Hz$^2$/Hz corresponds to an \emph{intrinsic} Lorentzian FWHM line width of $\Delta\nu_{\mathrm{l}}=\pi N_{\mathrm {w}}\simeq 200$~kHz. For frequencies greater than 500~kHz, the frequency noise PSD falls off rapidly due to the limited bandwidth of the photoconductive MCT detector. To ensure the reliability and assess the reproducibility of our results, we have repeated the experiment over several months using both the P(9) N$_2$O absorption line at 581.24~cm$^{-1}$ and the P(8) line at 582.1~cm$^{-1}$. Measurements were performed using injection currents ranging from 500 to 570~mA and laser temperatures from 246 K to 263 K. In all cases, the frequency noise spectra consistently showed the same trend, in particular the clearly visible white noise plateau above a few tens of kilohertz, confirming the intrinsic nature of these noise features. A second CW-DFB QCL of identical design, emitting at a nearby wavelength was used and shown to have the same frequency noise characteristics.

\begin{figure}[h]
\centering \includegraphics[scale=0.30]{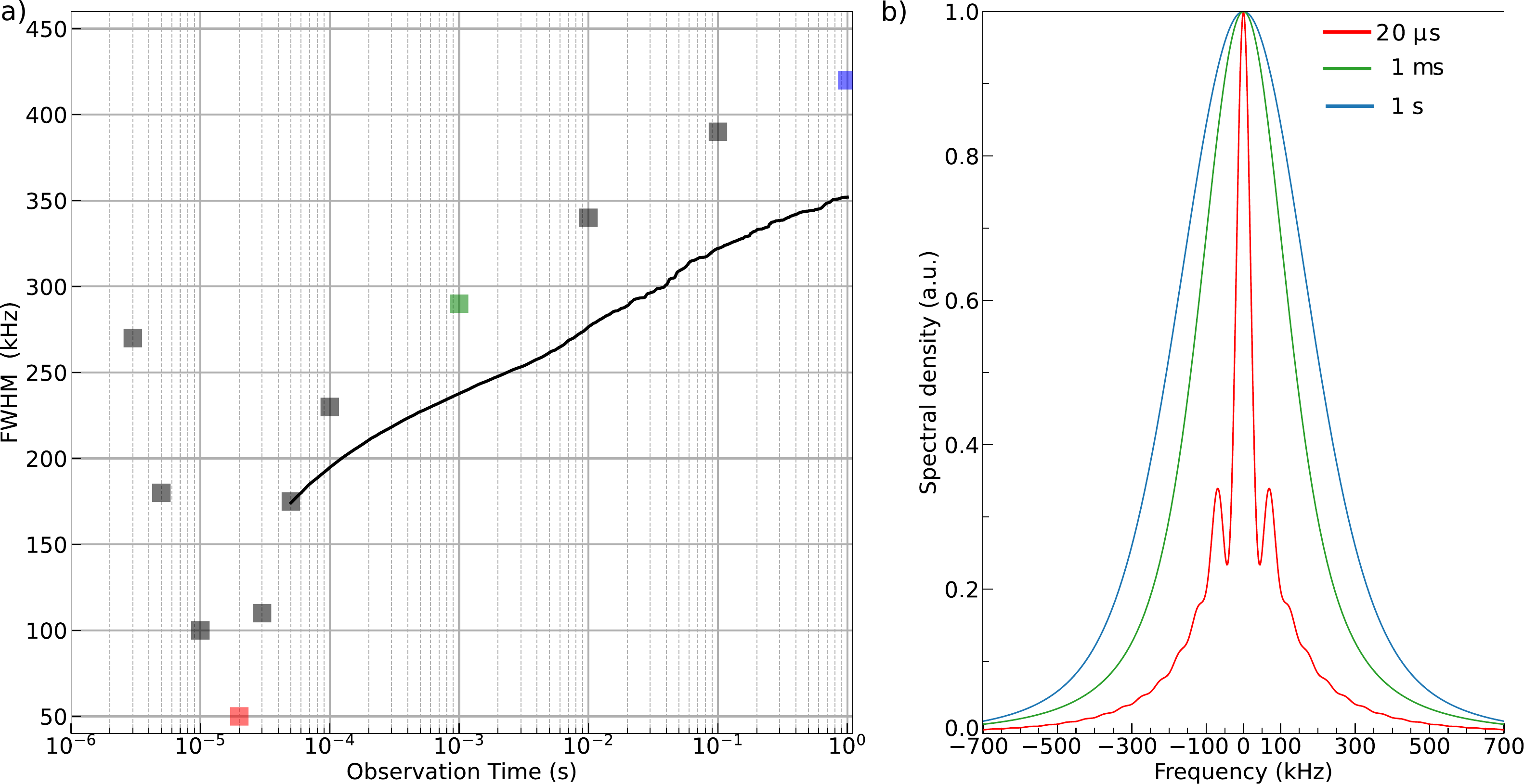} 
\caption{\label{fig:Linewidth} a): FWHM as a function of integration time. Square points: FWHM inferred from a full line shape reconstruction based on the measured frequency noise PSD (red curve in Figure~\ref{fig:NoisePSD}). Black line: FWHM determined using the $\beta$-separation line method. b): Corresponding QCL line shapes for three particular integration times: $1$~s (blue), $1$~ms (green) and 20~\textmu s (red). The same color code is used for the associated FWHM line widths in the a) panel.} \end{figure}

The \emph{real} QCL lineshape is broadened by flicker noise and its width thus depends on the observation time. Both can be determined using the frequency noise PSD. The FWHM is calculated to a good approximation using the $\beta$-separation line method described by Di Domenico \emph{et al.}~\cite{didomenico2010}. The $\beta$-separation line, shown as a green line in Figure~\ref{fig:NoisePSD}, is defined by $ 8 \ln(2) f / \pi^2$. The line separates the PSD into two regions: (i) below the cut-off defined by the crossing point between the line and the QCL frequency noise PSD ($f_{\mathrm{c}}\sim100$~kHz), and (ii) above $f_{\mathrm{c}}$. Region (i) is characterized by slow frequency modulation, where the noise has a high modulation index and thus contributes to the laser width. Region (ii) is characterized by fast frequency modulation, and thus low modulation index frequency fluctuations which contribute only to the wings of the line shape. These latter are consequently discarded in the estimation of the laser width. The estimated FWHM line width is given by $\sqrt{8\ln(2)A}$, with $A$ the area under the PSD in region (i). Figure~\ref{fig:Linewidth}a (black line) shows its evolution with the integration time $T_\textrm{int}$ (corresponding to the inverse of the lowest Fourier frequency considered). It reaches 350~kHz at $T_\textrm{int}=1$~s. It is not reported for small integration times $T_\textrm{int}<5/f_c=50$~\textmu s as the $\beta$-separation line method discards high frequency noise components and therefore fails to estimate correctly the line width for $T_\textrm{int}\times f_c<5$~\cite{didomenico2010}. To overcome this, we have calculated the \SMU laser line shape from the frequency noise PSD following reference \cite{elliott1982} and accounting for the integration time \cite{bishof2013,sow2014} (see \textbf{Figure~\ref{fig:Linewidth}} b)). The corresponding FWHM line widths are reported as square points on Figure~\ref{fig:Linewidth}. They are in moderate agreement with the $\beta$-separation line approximation. The latter is known to under-estimate the line width by $\sim$ 10\%  for pure $1/f$ noise~\cite{didomenico2010}. A larger $\sim20$\% disagreement is observed in our case, which we attribute to the contribution from the white noise plateau. The calculated line width decreases with integration time, reaching a minimum of 49~kHz when $T_{\rm int} = 20$~$\mu$s. For shorter $T_\textrm{int}$, the line width is Fourier limited by the measurement time and therefore increases again.

The theoretical framework reported by Yamanishi \emph{et al.}~\cite{yamanishi2007} provides a recipe for calculating the FWHM \emph{intrinsic} Lorentzian line width of a 3-level QCL. We use this formalism to calculate the expected \emph{intrinsic} FWHM of the \SMU laser, and we find it to be of the order of $\Delta\nu_{\mathrm{l,th}}\simeq 1 $~kHzunder our operating conditions. This is two to three orders of magnitude narrower than our experimental estimate (see above). Details of our calculations, including a comparison with other QCLs found in the literature, are presented in the Supplementary Materials. We are unable to explain the discrepancy between the measured noise plateau and the theoretical predictions, especially since the intrinsic line widths measured and inferred for QCLs at 4.3 and 10.6~\textmu m so far agreed with theoretical calculations, with reported values of a few $100$~Hz~\cite{bartalini2010,bartalini2011,sow2014,chomet_highly_2023}. However, the laser system discussed here operates at a significantly longer wavelength, and we are able to point out an essential difference compared to shorter wavelength QCLs which could contribute to the excess frequency noise in this wavelength regime: the gain coefficient is much larger in our \SMU source. This increased gain arises from a larger oscillator strength and narrower intersubband transitions. We estimate that the gain differs by approximately a factor of 30 between the 17~\textmu m laser and the 4.3~\textmu m lasers described in references \cite{bartalini2010,bartalini2011}. As a result, the population inversion  required to reach the threshold gain is very small, even smaller than the thermal population in the lower level of the laser transitions (level 2 in the Supplementary Materials). Fluctuations in this thermal population could be a source of additional noise not accounted for in the model described in reference \cite{yamanishi2007} used for the theoretical prediction.
The theoretical framework reported by Yamanishi \emph{et al.}~\cite{yamanishi2007} provides a recipe for calculating the FWHM \emph{intrinsic} Lorentzian line width of a 3-level QCL. We use this formalism to calculate the expected \emph{intrinsic} FWHM of the \SMU laser, and we find it to be of the order of $\Delta\nu_{\mathrm{l,th}}\simeq 1 $~kHzunder our operating conditions. This is two to three orders of magnitude narrower than our experimental estimate (see above). Details of our calculations, including a comparison with other QCLs found in the literature, are presented in the Supplementary Materials. We are unable to explain the discrepancy between the measured noise plateau and the theoretical predictions, especially since the intrinsic line widths measured and inferred for QCLs at 4.3 and 10.6~\textmu m so far agreed with theoretical calculations, with reported values of a few $100$~Hz~\cite{bartalini2010,bartalini2011,sow2014,chomet_highly_2023}. However, the laser system discussed here operates at a significantly longer wavelength, and we are able to point out an essential difference compared to shorter wavelength QCLs which could contribute to the excess frequency noise in this wavelength regime: the gain coefficient is much larger in our \SMU source. This increased gain arises from a larger oscillator strength and narrower intersubband transitions. We estimate that the gain differs by approximately a factor of 30 between the 17~\textmu m laser and the 4.3~\textmu m lasers described in references \cite{bartalini2010,bartalini2011}. As a result, the population inversion  required to reach the threshold gain is very small, even smaller than the thermal population in the lower level of the laser transitions (level 2 in the Supplementary Materials). Fluctuations in this thermal population could be a source of additional noise not accounted for in the model described in reference \cite{yamanishi2007} used for the theoretical prediction.

\section{Conclusion}

In conclusion, we have performed absorption spectroscopy of \ce{N2O} to demonstrate the spectroscopic capabilities of a new near-room-temperature CW DFB QCL operating at \SMU. This has allowed us to characterize the frequency noise of the laser and to measure its line width, and to bring to light puzzling discrepancies with the currently reported theoretical understanding of QCL frequency noise. This laser operates in a spectral region that is poorly covered by existing lasers, and its development opens new opportunities in atmospheric sensing and chemical detection, as well as in precise spectroscopic tests of fundamental physics. This specific narrow-line width laser source emitting at \SMU offers perspectives for the manipulation of bismuth spin states in silicon for solid-state quantum technology and atomic clock applications~\cite{saeedi_optical_2015} or could be exploited as a local oscillator in a heterodyne detector for astrophysical molecular detection around 17~\textmu m~\cite{bourdarot2020,bourdarot2021}. We have also recently used it for wavelength modulation laser spectroscopy of \ce{N2O} and demonstrated its potential for building accurate rovibrational spectroscopic models~\cite{wang_wavelength_2025}. The present work is also the first step towards frequency stabilization of this source for subsequent precise spectroscopy in this region. For applications in frequency metrology, the frequency of the QCL can be stabilized using one of several techniques. For example, the frequency can be stabilized to a molecular absorption line, e.g. in \ce{N2O}. Alternatively, it could be stabilized to an optical cavity that bridges the gap from 17 \textmu m to another wavelength where narrow-linewidth, stabilized lasers are readily available, e.g. near 1 \textmu m. For the ultimate performance, the laser should be stabilized to a frequency comb using non-linear optics as already demonstrated at 10 \textmu m and below ~\cite{hansen_quantum_2015,argence2015,insero2017,santagata2019}. Extending frequency metrology techniques to QCLs operating beyond 15 \textmu m is particularly compelling for high-resolution spectroscopy of large molecules. In this low energy spectral region, rovibrational coupling mechanisms which can result in severe spectral blurring are significantly reduced. This opens up the possibility of using increasingly complex polyatomic molecules to perform tests of fundamental physics, such as to measure the energy difference between enantiomers in heavy chiral molecules, a signature of parity violation induced by the weak interaction~\cite{cournol2019,fiechter_towards_2022}, and a sensitive probe of dark matter~\cite{gaul_chiral_2020}. Longer wavelength QCLs are also necessary to develop frequency standards in the MIR based on ultra-cold diatomic molecules. The 17.2~\textmu m QCL used in the present study has been designed to coincide with the fundamental vibrational frequency of CaF~\cite{kaledin1999}, one of the few molecules that has been laser cooled down to the microkelvin range~\cite{truppe2017}. A clock based on the fundamental vibrational transition of ultra-cold CaF molecules confined in an optical lattice is currently under construction~\cite{barontini_measuring_2022}. It is expected to have a line width below 10 Hz, a stability of $2\times10^{-15}$ at 1~s, and the potential to measure the stability of the electron-to-proton mass ratio to a fractional precision better than $10^{-17}$ per year.

\section*{Supporting information}

\begin{figure}[h]
\centering \includegraphics[scale=0.48]{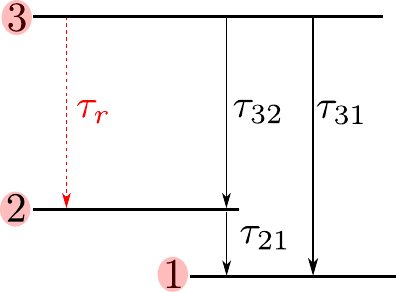} 
\caption{\label{fig:Levels} Schematics of the quantum levels of a given potential well for a 3-level QCL. The radiative relaxation between level 3 and 2 producing photons is represented by the red dashed arrow and is characterized by a relaxation time $\tau_{\mathrm{r}}$. The other displayed relaxation paths are non-radiative.} \end{figure}

In this supplementary material, we derive the calculation of the theoretical intrinsic line width $\Delta\nu_{\mathrm{l,th}}$ of the \SMU laser, following the Yamanishi method \cite{yamanishi2007} for a 3-level QCL, as depicted in Fig.\ref{fig:Levels}. Only the parameters relevant to this discussion are presented here; additional details on the QCL design can be found in reference~\cite{nguyen2019}. Using the notations of Yamanishi \emph{et al.}~\cite{yamanishi2007}, the \SMU QCL is characterized by the following geometric and optical properties: 

\begin{itemize}
     \item a quantum-confinement structure thickness $N\times t_{\mathrm{qc}}=5.2$~\textmu m in the $z$-direction perpendicular to the plane of the $N=55$ cascade stages, each of thickness $t_{\mathrm{qc}}$; a lateral width of the quantum-confinement structure $w_{\mathrm{qc}}=14$~\textmu m in the $y$-direction; a refractive index of the waveguide active material of $n_{\mathrm{g}}=3.38$, and a cladding index $n_{\mathrm{c}}=3.0$;
     \item a cavity of length $L_{\mathrm{c}}=3$~mm along the $x$-direction, exhibiting a single guided TM$_{00}$ mode of effective refractive index  $n_{\mathrm{eff,0,0}}=3.34$ and group index $n_{\mathrm{g,0,0}}=3.77$; the fractional energy associated with the $z$-component of the electric field of the guided mode, $\varGamma_{\mathrm{TM,0,0}}$ (see equation (A9d) in~\cite{yamanishi2007}), is unity in our single transverse mode case; the optical confinement of the mode is such that $\varGamma_{\mathrm{Conf,y0}}=1$ and $\varGamma_{\mathrm{Conf,z0}}=0.64$ (see respectively equations (A10b) and (A16b) in~\cite{yamanishi2007}); 
     \item the overall cavity losses are $\alpha_{\mathrm{tot}}=3500$~m$^{-1}$ (the sum of internal, $\alpha_{\mathrm{int}}$, and mirror, $\alpha_{\mathrm{m}}$, losses, as denoted in~\cite{yamanishi2007}), which gives a photon decay rate of $\gamma = \dfrac{\alpha_{\mathrm{tot}}c}{n_{\mathrm{g,0,0}}}=0.28$~THz, with $c$, the speed of light in vacuum;
     \item a $z$-oriented dipole moment of $\langle\phi_3|z|\phi_2\rangle$ of $6.5$~nm;
     \item a full-width-at-half-maximum $\hslash\varGamma=5$~meV of the lineshape function (supposed to be Lorentzian here) associated to the intersubband transition from state 3 to state 2 in Fig.~\ref{fig:Levels}.
\end{itemize}     

As given in references~\cite{bartalini2010} (equation 2) and \cite{yamanishi2007} (equation 16a), the intrinsic line width is:

\begin{equation}
\Delta\nu_{\mathrm{l,th}}=\frac{\gamma\beta_{\mathrm{eff}}}{4\pi(1-\epsilon)}\bigg[\frac{1}{I_0/I_{\mathrm{th}}-1}+\epsilon\bigg](1+\alpha_{\mathrm{c}}^2),
\label{eq:Width}
\end{equation}

\noindent with:
\begin{itemize}
     \item  $\epsilon=\frac{\tau_{21}\tau_{31}}{\eta\tau_{\mathrm{t}}(\tau_{21}+\tau_{31})}$, a parameter which depends on the relaxation times of the various levels, see Fig.\ref{fig:Levels}, $\tau_{\mathrm{t}}$ being the total relaxation time of the upper level : $\tau_{\mathrm{t}}=(1/\tau_{\mathrm{r}}+1/\tau_{32}+1/\tau_{31})^{-1}$, with $\tau_{\mathrm{r}}$ radiative (spontaneous emission) and $\tau_{32}$, $\tau_{31}$ non-radiative relaxation times. Finally, $\eta$ represents the injection efficiency  of the charges in level 3;
    \item  $\alpha_{\mathrm{c}}$, the Henry line width enhancement factor,  typically small for QCLs, usually ranging from 0 to 2 ~\cite{faist1994,vonstaden2006,kumazaki2008,jumpertz2016,franckie2023};
    \item $I_{\mathrm{th}}$  and $I_0$ the  lasing threshold and injected current respectively;
    \item $\gamma$ the photon decay rate;
    \item $\beta_{\mathrm{eff}}$ the \emph{effective coupling efficiency} (as denoted in~\cite{yamanishi2007}) of the emission from level 3 in Fig.\ref{fig:Levels}. It is the ratio of the spontaneous emission rate coupled into the lasing mode -- \emph{i.e.} $\beta/\tau_{\mathrm{r}}$, with $1/\tau_{\mathrm{r}}$, the spontaneous emission rate defined above, and $\beta$, the spontaneous emission coupling efficiency to the lasing mode -- to the total relaxation rate from the upper level $1/\tau_{\mathrm{t}}$ (see above). We thus have: $\beta_{\mathrm{eff}}=\dfrac{\beta/\tau_{\mathrm{r}}}{1/\tau_{\mathrm{t}}}$.

\end{itemize}

The effective coupling $\beta_{\mathrm{eff}}$ is typically small for QCLs due to very efficient non-radiative processes ($1/\tau_{\mathrm{r}}\ll1/\tau_{\mathrm{t}}$). This in turn explains the narrow intrinsic line width of QCLs due to a reduction of the noise associated with spontaneous emission. The order of magnitude of $\Delta\nu_{\mathrm{l,th}}$ is largely determined by the product $\gamma\beta_{\mathrm{eff}}$. We now focus on calculating the effective coupling efficiency $\beta_{\mathrm{eff}}$ for the specific case of the single transverse mode \SMU QCL. We start by determining the spontaneous emission coupling efficiency to the lasing mode $\beta=\beta_{\mathrm{guide}}\times\beta_{\mathrm{l}}$ which can be expressed as the product of $\beta_{\mathrm{guide}}$, the coupling efficiency of spontaneous emission to the transverse guided mode from the total spontaneous emission, and $\beta_{\mathrm{l}}$, the coupling efficiency to a given longitudinal mode $\mathrm{l}$ from the specific transverse guided mode~\cite{yamanishi2007}, with: 
\begin{itemize} 
\item $\beta_{\mathrm{guide}}=\dfrac{1/\tau_{\mathrm{rg}}}{1/\tau_{\mathrm{r}}}=\dfrac{1/\tau_{\mathrm{rg}}}{1/\tau_{\mathrm{rfp}}+1/\tau_{\mathrm{rg}}}$ and $\beta_{\mathrm{l}}=\dfrac{ \hslash c }{n_{\mathrm{g,0,0}}L_c}\dfrac{2}{\hslash\varGamma}$
\item $\tau_{\mathrm{rfp}}$, the relaxation time of the spontaneous emission coupled to the free-space continuum modes defined by (see equation (A6b) in~\cite{yamanishi2007}):
\begin{equation}
\dfrac{1}{\tau_{\mathrm{rfp}}}=\dfrac{4\pi^2 e^2n_g}{\epsilon_0\hslash\lambda_0^3}|\langle\phi_3|z|\phi_2\rangle|^2\Big[1-\frac{3}{2}\Big(1-(\frac{n_c}{n_g})^2\Big)^{1/2}+\frac{1}{2}\Big(1-(\frac{n_c}{n_g})^2\Big)^{3/2}\Big];
\end{equation}
according to which we estimate $\tau_{\mathrm{rfp}}=92$~ns;
\item $\tau_{\mathrm{rg}}$, the relaxation time of spontaneous emission coupled to the guided mode defined by (see equation (A10a) in~\cite{yamanishi2007}):
\begin{equation}
\dfrac{1}{\tau_{\mathrm{rg}}}=\dfrac{2\pi e^2} {\epsilon_0\hslash\lambda_0w_{\mathrm{qc}}}|\langle\phi_3|z|\phi_2\rangle|^2\dfrac{n_{g,0,0}}{n^2_{\mathrm{eff,0,0}}}\varGamma_{\mathrm{TM,0,0}}\varGamma_{\mathrm{Conf,y0}}\dfrac{\varGamma_{\mathrm{Conf,z0}}}{N t_{\mathrm{qc}}};
\end{equation}
according to which we estimate $\tau_{\mathrm{rg}}=793$~ns;
\item a total radiative lifetime $\tau_{\mathrm{r}}=\Big(1/\tau_{\mathrm{rg}}+1/\tau_{\mathrm{rfp}}\Big)^{-1}=82$~ns.
\end{itemize}
this leads to a coupling efficiency of spontaneous emission to the transverse guided mode of $\beta_{\mathrm{guide}}=0.10$ and a coupling efficiency of spontaneous emission to a single longitudinal mode from the specific transverse guided mode of $\beta_{\mathrm{l}}=0.0069$, and thus a coupling efficiency of spontaneous emission into the lasing mode of $\beta=\beta_{\mathrm{l}}\times\beta_{\mathrm{guide}}=7.2\times10^{-4}$.
\\

Table \ref{tab:Geometry} summarizes the values of the various parameters of equation \ref{eq:Width} for the \SMU laser. The numerical values characterizing the \SMU QCL reported in the upper part of Table \ref{tab:Geometry} as well as above come from simulations. They lead to fairly good agreement with the main measured QCL features: threshold, gain, wavelength, geometry, temperature dependence ... We also report in Table \ref{tab:Geometry} the corresponding values for a $4.33$~\textmu m DFB QCL cooled to liquid-nitrogen temperatures~\cite{bartalini2010}  and a $4.36$~\textmu m QCL working at room temperature~\cite{bartalini2011}.

Note that combining the formulas of $\beta_{\mathrm{eff}}$, $\beta$ and $\beta_{\mathrm{guide}}$ given above leads to $\beta_{\mathrm{eff}}=\dfrac{\beta_{\mathrm{l}}/\tau_{\mathrm{rg}}}{1/\tau_{\mathrm{t}}}$. We then see that $\beta_{\mathrm{eff}}$ is quasi-independent of $\tau_{\mathrm{r}}$ which only appears in $\tau_{\mathrm{t}}$ in which it makes a negligible contribution in comparison to $\tau_{\mathrm{31}}$ and $\tau_{\mathrm{32}}$. In fact, $\tau_{\mathrm{r}}$ and in turn $\tau_{\mathrm{rfp}}$ are not really needed to estimate $\beta_{\mathrm{eff}}$.

Finally, we infer an effective coupling efficiency $\beta_{\mathrm{eff}}$ of $1.9\times10^{-9}$ for the \SMU QCL, and deduce from equation \ref{eq:Width} a theoretical intrinsic line width $\Delta\nu_{\mathrm{l,th}}$ between 340~Hz and 1700~Hz considering typical line width enhancement factor values, for an injected current of $I=570$~mA and a lasing threshold of $I_{\mathrm{th}}=480$~mA at a temperature of $T=258$~K, corresponding to our experimental conditions. The corresponding values are also given for the $\sim4.3$~\textmu m QCLs found in the literature for comparison.

\renewcommand{\arraystretch}{1.4}
\begin{table*}[h]
\centering
\begin{tabular}{| c | c | c | c |}
  \hline
  & Reference \cite{bartalini2010} & Reference \cite{bartalini2011} & \SMU QCL \\ 
  \hline
  \hline
 $\tau_{21}$ (s) & $0.25 \times 10^{-12}$ & $0.15 \times 10^{-12}$ & $0.073 \times 10^{-12}$\\
  \hline
 $\tau_{31}$ (s)  & $2 \times 10^{-12}$ & $1.79 \times 10^{-12}$ & $0.29 \times 10^{-12}$ \\
  \hline  
 $\tau_{32}$ (s) & $3.4 \times 10^{-12}$ & unknown &  $0.92 \times 10^{-12}$\\
  \hline
 $\tau_{r}$ (s) & $7.5 \times 10^{-9}$ &  $10 \times 10^{-9}$ & $82 \times 10^{-9}$\\
  \hline
 $\tau_{\mathrm{t}}$ (s) & $1.26 \times 10^{-12}$ & $1 \times 10^{-12}$ & $0.22 \times 10^{-12}$ \\
  \hline
 $\eta$ & $0.7$ & $0.7$ & $0.9$ \\
 \hline
  \hline
 $\gamma$ (Hz) & $1.2 \times 10^{11}$ & $1.2 \times 10^{11}$ & $2.8 \times 10^{11}$ \\
  \hline
 $\beta_{\mathrm{eff}}$ & $1.6 \times 10^{-8}$ & $ 5\times 10^{-9}$ & $1.9\times10^{-9}$ \\
 \hline
 $\epsilon$ & $0.25$ & $0.2$ & $0.29$ \\
  \hline
 $I/I_{\mathrm{th}}$ & $1.54$ & $1.15$  & $1.19$ \\
 \hline
  \hline
 $\bm{\Delta \nu_{\mathrm{l,th}}}$ \textbf{(Hz)} & $\mathbf{510}$  & $\mathbf{340}$  & $\mathbf{340-1700}$ \\
  \hline
\end{tabular} 
   \caption{Values of the parameters to calculate the 
   intrinsic QCL line width ${\Delta \nu_\mathrm{{l,th}}}$ as defined in equation~\ref{eq:Width}, and a comparison between the \SMU QCLs and QCLs presented in references \cite{bartalini2010,bartalini2011}. 
   A value between 340 and 1700~Hz for  ${\Delta \nu_\mathrm{{l,th}}}$ is found for the \SMU QCL depending on the Henry line width enhancement factor : $\alpha_{\mathrm{c}}$ ranging from 0 to 2.}
  \label{tab:Geometry}
\end{table*}

\section*{Acknowledgements}

TEW acknowledges funding from the Royal Society International Exchanges Scheme (grant IES\textbackslash R3\textbackslash 183175), the Imperial College European Partners Fund and the Université Sorbonne Paris Nord Visiting Fellow Fund. MRT acknowledges the support from the UK STFC (ST/T006234/1 and ST/W006197/1) for this research. This work has been supported (i) the People Programme (Marie Curie Actions) of the European Union’s Seventh Framework Programme (FP7/2007-2013) under REA grant agreement n. PCOFUND-GA-2013-609102, through the PRESTIGE programme coordinated by Campus France; (ii) Region Ile-de-France in the framework of DIM SIRTEQ and DIM QuanTiP; (iii) the Imperial College London -- CNRS 2021 PhD joint programme; (iv) CNRS; (v) Université Sorbonne Paris Nord. This work was part of 23FUN04 COMOMET that has received funding from the European Partnership on Metrology, co-financed by the European Union’s Horizon Europe Research and Innovation Programme and from by the Participating States, funder ID: 10.13039/100019599.

\bibliographystyle{unsrt}
\bibliography{17micronLaserBib.bib}

\end{document}